\documentclass[prl,twocolumn,showpacs,reprint,superscriptaddress]{revtex4}
\usepackage{amssymb}
\usepackage{amsmath}
\usepackage[T1]{fontenc}
\usepackage{amsfonts}
\usepackage{graphicx}
\usepackage{subfigure}
\usepackage{dcolumn}
\usepackage{bm}
\usepackage{hyperref}
\usepackage{empheq}
\usepackage{natbib}
\usepackage{color}

\begin{document}
\preprint{Version0.6}

\title{Identifying polymer states by machine learning}

\author{Qianshi Wei}
\affiliation{Department of Physics and Astronomy, University of Waterloo, Waterloo N2L 3G1, Canada}

\author{Roger G. Melko}
\affiliation{Department of Physics and Astronomy, University of Waterloo, Waterloo N2L 3G1, Canada}
\affiliation{Perimeter Institute for Theoretical Physics, Waterloo, Ontario N2L 2Y5, Canada}

\author{Jeff Z. Y. Chen\footnote{jeffchen@uwaterloo.ca}}
\affiliation{Department of Physics and Astronomy, University of Waterloo, Waterloo N2L 3G1, Canada}

\date{\today}

\begin{abstract}
The ability of a feed-forward neural network to learn and classify different states of polymer configurations is systematically explored. Performing numerical experiments, we find that a simple network model can,  after adequate training, recognize multiple structures, including gas-like coil, liquid-like globular, and crystalline anti-Mackay and Mackay structures. The network can be trained to identify the transition points between various states, which compare well with those identified by independent specific-heat calculations. Our study demonstrates that neural network provides an unconventional tool to study the phase transitions in polymeric systems.
\end{abstract}

\pacs{61.25.H-, 05.70.Fh, 05.10.-a, 36.40.Ei}

\maketitle

{\it Introduction.--} 
Machine learning is an active research branch of computer science,
which has vastly matured in recent years \cite{frawley1992knowledge, watkin1993statistical, witten2005data, michalski2013machine, murphy2012machine, jordan2015machine}. Artificial neural networks (NN) are a computational implementation of machine learning that
have demonstrated surprising capability in recognizing patterns of enormous complexity, after appropriately trained by human or self-trained through learning mechanisms \cite{fukushima1988neocognitron, seung1992statistical, parekh2000constructive, jain2000statistical, davatzikos2005classifying, bishop2006pattern, schmidhuber2015deep}.
Originally motivated by the desire to establish an algorithmic model of the neuronal configuration of a mammalian brain, one finds that an artificial NN with 
a very simple underlying structure can already successfully perform many complex tasks. For example, simple NN models have shown transformative success in hand-writing and speech recognitions~\cite{lecun1995learning,nielsen2015neural,lippmann1989review,hinton2012deep}.

The application of machine learning in computer simulations of molecular systems 
has only recently emerged.
Machine learning has been proposed as a method for obtaining approximate solutions for the equations
of density functional theory \cite{Burke}; for constructing force-fields for molecular dynamics \cite{MD1}; and for providing an inexpensive
impurity solver for dynamical mean-field theory~\cite{arsenault2014machine}.
More recently, 
a simple supervised NN was implemented
to classify phases in a two-dimensional (2D) spin model and to identify critical temperatures
between various phases directly from configurations produced by Monte Carlo (MC) simulations~\cite{carrasquilla2016machine}.
This comes on the crest of a wave of work exploring the hybridization of MC and machine learning~\cite{Torlai2016,Huang_Wang,LiangFu,Tamblyn},
and the deeper connections between techniques such as deep learning to theoretical concepts familiar from
statistical mechanics, such as the renormalization group~\cite{mehta2014exact, RG2}.

Here, we demonstrate the capability of a standard NN model in recognizing various configurations produced from MC simulations of polymer models, identifying disordered, partially-ordered, and ordered states such as coil, liquid-like globular, and two low-energy crystalline structures.
The physical system we discuss has a long history, hence is used 
as a typical example of a classical system displaying phase transitions that can be conveniently studied by using a hybridization of conventional simulation and machine learning techniques.

\begin{figure}[!t]
        \centering
        \includegraphics[width=0.8\columnwidth]{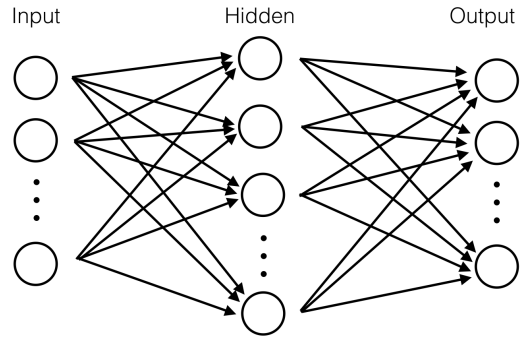}
        \caption{\label{ANN} Sketch of a fully connected feed-forward artificial neural network. The circles represent neuron nodes connected layer by layer via various weight and activation parameters.}
\end{figure}

{\it The feed-forward neural network.--}
A standard feed-forward NN for supervised learning consists of three layers of neurons or ``nodes'', input (i), hidden (h), and output (o), each containing $N_{\rm i}$, $N_{\rm h}$,  $N_{\rm o}$ neurons [see Fig. \ref{ANN}].
These nodes are connected thourgh edges (representing model weights), forming a fully-connected graph between (but not
among) each neuron layer.
In a typical image recognition application, a 2D picture is discretized into $N_{\rm i}$ pixels and the values of 
pixel's intensity
are then fed into the input layer for either training or testing purposes. The configurations of a 2D Ising model, where $N_{\rm i}$ spins have binary values, share some similarity to the standard image recognition problem, as each configuration is a ``snapshot'' of a 2D pattern of pixels.  However, in a simple feed forward NN, the details of this dimension and locality are lost as pixels are simply sent into input nodes as a one-dimensional vector~\cite{carrasquilla2016machine}.

A simulated, three-dimensional (3D) polymer configuration is mathematically represented by its $3N$ spatial coordinates of the $N$ connected monomers.  In order to proceed, one must decide how this information is fed into the NN.
One option is to naively adopt the image recognition idea above, by digitizing the 3D space into a grid and assigning a value of ``occupied'' or ``unoccupied'' to a particular grid point, describing the occupancy of monomers. 
The entire pattern would then be used for analysis. Alternatively, 
here we analyze a polymer configuration by directly feeding $3N$ coordinates into $N_{\rm i} = 3N $ input nodes.  Then, to implement supervised learning, the 
NN is trained to analyze the relationship between the monomer position vector and a configurational pattern corresponding to the ``label'' in the output layer.  In total $N_{\rm h}=100$ neurons are used and the number of the output nodes, $N_{\rm o}$ is set to 2 or 3 depending on the physical problems described below. As a technical note, during the training session, we adopted  cross entropy as the cost function and used 50\% dropout rate in determination of the parameters related to the hidden layer, which is a regularization method to avoid overfitting~\cite{Srivastava}.

\begin{figure}
        \includegraphics[width=1\columnwidth]{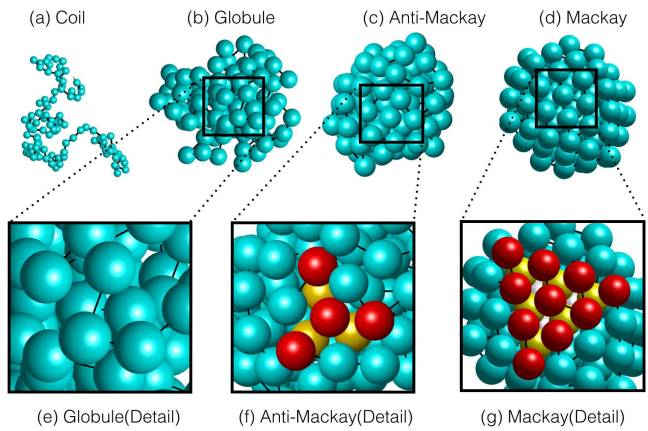}
        \caption{\label{States} Typical configurations of a polymer in 
        (a) coil, (b) globular, (c) anti-Mackay, and (d) Mackay states, as the temperature is lowered. The $N=102$ monomers are represented by blue spheres, except for those in the last two plots. At the liquid-like globular state, the monomer positions are disordered, shown in (e). Both anti-Mackay and Mackay states are crystalline, differ from each other by the monomer stacking symmetries, demonstrated by the red and yellow spheres in (f) and (g).}
\end{figure}

{\it Polymer states.--} Two polymer models are used here, both consisting of $N=102$ connected monomers 
interacting through a typical short-range hard-core repulsion and a slightly-longer-range attraction. One of the polymer models is known to exhibit four different states illustrated in Fig.~\ref{States}, separated by phase transitions 
of finite-size characteristics  \cite{Schnabel2009,seaton2010collapse,Backmann2011}.
Within the first model, the bonded monomers interact with each other by a spring potential, identical to the one used in the Gaussian model (GSM) with a Kuhn length $a$~\cite{DoiandEdwards}; two non-bonded monomers interact with each other by a square-well potential: below a square distance of $0.81a^2$ the monomers repel through the excluded-volume interaction, and between 
$0.81a^2$ and $2a^2$ the monomers experience an attraction of magnitude $-\epsilon$. Within the second model, the bonded monomers are modeled by a particular implementation of the Finitely Extensible Nonlinear Elastic (FENE) model, and the 
monomers interact with each other in the form of the Lennard-Jones (LJ) potential with a potential well-depth $-\epsilon$. The selection of FENE and LJ functions exactly matches those used in Ref. \cite{Backmann2011}. In this Letter we refer to the first model as GSM and second as FENE for simplicity. More details are defined in the supplemental material~\cite{SupplementalMaterial}.

\begin{figure*}
        \centering	
        \includegraphics[width=2.00\columnwidth]{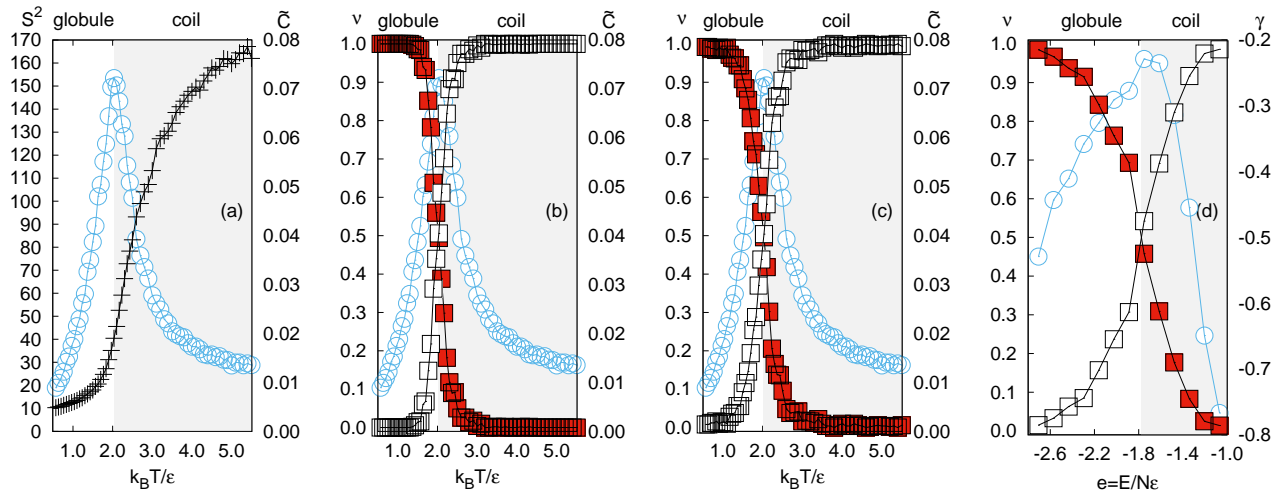}
        \caption{\label{Fig3}(a) Reduced mean-square radius of gyration (plus symbols, to the left scale) and specific heat (circles, to the right scale) as functions of the reduced temperature $k_{\rm B}T/\epsilon$, determined from the MC simulations of GSM for $N=102$, (b) the NN outputs of test-recognizing independent GSM configurations, (c) the  NN outputs of test-recognizing independent, {\emph {normalized}} GSM configurations, and (d) the NN outputs of test-recognizing normalized FENE configurations. The filled and open squares represent the mean $\nu$-values output from the globular and coil neurons, respectively. The circles in plot (d) represent the specific-heat-like $\gamma$ (to the right scale) determined for the FENE model from a Wang-Landau MC simulation. Error bars associated with all squares are smaller than the symbol size. }
\end{figure*}

{\it Coil-to-globule transition.--} MC simulations which incorporate the Boltzmann weight were used for GSM to produce  $5\times 10^3$ independent configurations at every specified temperature $k_{\rm B}T/\epsilon$ where $k_{\rm B}$ is the Boltzmann constant. Assessing the data shown in Fig. \ref{Fig3}(a) for both reduced mean-square radius of gyration, $S^2 \equiv \langle R_{\rm g}^{2} \rangle/a^2 $, and mean-square deviation of the total energy from its average (which is proportional to the specific heat), ${\tilde C} \equiv [\langle E_{\rm int}^{2} \rangle- \langle E_{\rm int}\rangle^{2}]/N^2\epsilon^{2} $, we observe that  a coil(C)-to-globule(G) phase transition takes place at $k_BT/\epsilon \approx 2.0 $, corresponding to the location of the peak  in  ${\tilde C}$.

To demonstrate the capability of NN in recognizing C and G, we performed three numerical experiments.
First, 
$3\times 10^3$ polymer configurations (
specified by the coordinates of  monomers after setting $a=1$) at every specified temperature within the ranges $[0.5, 1.5]$ and $[2.5, 3.5]$ 
were selected as training sets for the G and C states, respectively. The two normalized output neurons were designated as the G and C labels, which
during training were assigned to have values $\nu=1$ for the corresponding state, and $\nu=0$ otherwise.
No other information or estimators typical of MC simulations, such as
$S^2$ or $\tilde C$
were used in the training. Once the NN was adequately trained and all NN parameters were fixed, we input $500$ new configurations at every temperature between the range $[0.5, 5.5]$, which were not used in the training session, as the testing set. The averaged test values of the two output neurons, $\nu$,  are plotted in Fig \ref{Fig3}(b) forming two curves behind the filled and open squares. 
These curves cross each other, identifying a C-to-G transition at $k_BT/\epsilon = 2.03$, in agreement with the location of the $\tilde C$-peak, regardless of the fact that the NN model was trained in temperature ranges farther away from the transition point.

The C and G states have distinguishably different dimensions, represented by 
$S^2$. To show that NN is not simply recognizing 
C and G from their different overall sizes, in our second experiment we normalized all coordinates, of the training and testing sets, by a factor $1/S$. 
On average, the polymer configurations now have the same normalized radius of gyration $(=1)$, across the entire studied temperature range. The network was then trained in a similar manner described above, with the normalized coordinates. The quality of output neurons to indicate the C and G states for the testing data is equally good as in the previous case, shown in Fig.~\ref{Fig3}(c).

While these numerical experiments were performed by using the configurational data generated from GSM, we placed the network into the ultimate test in the third numerical experiment. This time, the NN parameters determined in the second experiment were retained and we asked the network to recognize the configurations generated from the FENE model, in which completely different potential functions were used than in the square-well GSM. Furthermore, instead of producing the configurations from the canonical ensemble where $k_BT/\epsilon$ is used as the system parameter, we generated configurations from the Wang-Landau (WL) algorithm for microcanonical ensembles, in which the total (reduced) energy per particle $e=E/N\epsilon$ is directly used as the system parameter. At each $e$-value illustrated in Fig. \ref{Fig3}(d), $1\times 10^{3}$ independent FENE configurations, normalized by their corresponding $S$,
were sent to the GSM-trained network for testing. The two recognition curves produced from the output neurons, shown in the figure as the underlying curves behind filled and open squares, predicate a C-to-G transition 
at $e  =-1.8$. This prediction can be independently verified by examining the $\gamma(e)$ curve, which is a specific-heat-type  measurement in reduced units defined in the microcanonical ensemble~\cite{Backmann2011}. The data represented by circles in Fig. \ref{Fig3}(d) was calculated based on the density of states determined from the WL algorithm; it shows a peak at the same location as the one successfully predicted by GSM-trained NN.

\begin{figure}[!t]
        \centering
        \includegraphics[width=0.95\columnwidth]{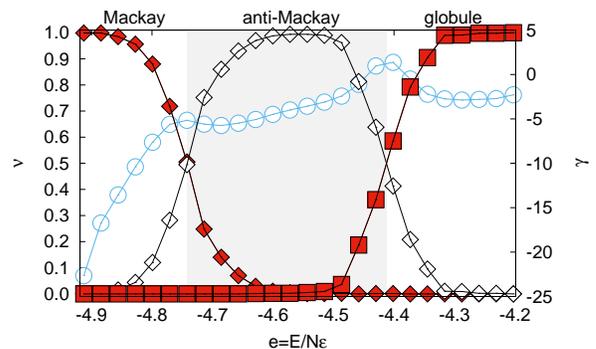}
        \caption{\label{les} Mean NN outputs $\nu$ (square for globule, open diamonds for anti-Mackay, and filled diamonds for Mackay) from the test samples, after the network is trained to recognizing these states in regimes where they are stable. In the background, the reduced specific heat-like $\gamma$ (circles, to the right scale) was independently produced from the MC simulations. 
        Error bars are smaller than the symbol size.  }
\end{figure}

{\it Low-energy polymer states.--} It is well-known that the FENE model exhibits three different states in the low-energy region, globule, anti-Mackey~(aM), and Mackey~(M), reported by careful MC studies utilizing the WL algorithm which uses $e$ as the system parameter~\cite{seaton2010collapse, Backmann2011}. One could convert all languages to a low-temperature description but we stay here with the low-energy description. These three structures have subtle structural differences which cannot be distinguished by direct visualization in Figs. \ref{States}(b), (c) and (d). In particular, the crystalline aM and M
differ delicately from each other by the way that monomers are stacked
[Figs. \ref{States}(f) and (g)].

As a final numerical experiment, we challenge the NN 
to recognize these three states, using three neuron nodes in the output layer, each assigned to recognize G, aM and M 
separately. The network was trained with FENE configurations in the energy range $e=[-4.3,-4.16]$, [-4.7,-4.5], and [-4.9,-4.8], where the G, aM and M structures, respectively, can be clearly defined. The training data contained 3$\times 10^3$ configurations at every energy bin.

After the network was adequately trained, it was tasked to recognize an independent set of test data covering the entire energy range in Fig \ref{les}. Over $10^3$ configuration samples were used at every energy bin for this purpose. The mean $\nu$-values of the test output, from the G, aM, and M nodes, are represented in Fig \ref{les} by filled squares, open diamonds, and filled diamonds. The intersections of the interpolated curves 
predict that G-to-aM and aM-to-M transitions 
take place at $e=-4.40$ and $e=-4.74$, respectively. These NN predicted transition points can be 
confirmed
by an examination of 
$\gamma$,
independently calculated from the WL MC simulations. From the $\gamma$-peaks 
in the plot, we determine that the G-to-aM and aM-to-M transition points are at $e=-4.40 \pm 0.03$ and $e=-4.74 \pm 0.03$ respectively, which agree 
with those from the NN predictions.

{\it In search of a phase transition.--} The above numerical experiments demonstrated the NN's versatility in recognizing polymer configurations and the usefulness of NN in determination of the transition points. There are two essential questions we have not adequately addressed. How does the NN predicted location of the transition depend on the range of used training data? Would the network mistakenly identify a phase transition for a system where a certain physical property smoothly crosses over from relatively large to small values without going through a real phase transition?

\begin{figure}[!t]
        \centering
        \includegraphics[width=1\columnwidth]{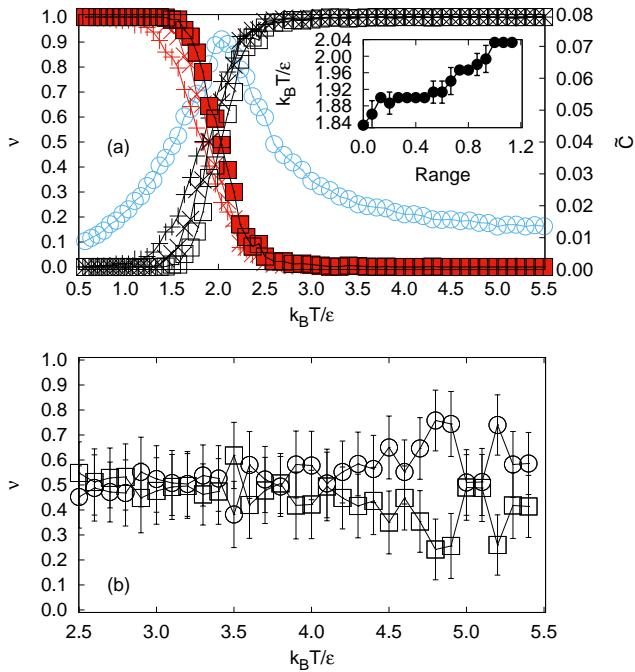}
        \caption{\label{compare}
        (a)  Average NN output $\nu$ from the globule and coil neurons on the testing configurations for NN models trained in various temperature ranges
        and (b) average NN output $\nu$ from ``phase-1''- (squares) and ``phase-2''-nodes (circles) on the test configurations in the temperature range $[2.5, 5.5]$ for NN models trained by using low- and high-temperature 
        samples. The inset in plot (a) is the NN-predicted transition temperature as a function of the training temperature-range used. Error bars in (a) are smaller than the symbol size. The plus, cross, and square symbols represent the mean $\nu$ from the two output neurons on test samples, of NN models initially trained in the temperature range,  ${\rm Range}=0.00, 0.53$, and $1.20$, respectively. The circles in (a) represent the same data in Fig. \ref{Fig3}(a). The temperature range in (b) contains no actual phase-transition point.
        }
\end{figure}

To answer the first question we return to the GSM MC data 
over a wide range of the reduced temperature, $[0.5,3.5]$. We conducted a series of 18 independent training sessions, each assigned a different training temperature range, away from the C-to-G transition point. The first session treated MC configurations produced from the system at two temperature values, one on the extreme left and the other extreme right, where the C-to-G transition sits in the middle. In the subsequent sessions the training temperature ranges expanded from the extreme left to the center and extreme right to the center, incrementally. These trained networks were then put into test, by inputting independent configurations over the extended $[0.5,5.5]$-range. The mean output $\nu$-values from the C and G nodes are illustrated in Fig \ref{compare}(a) for a few selected sessions. 
As the training range expands the NN-predicted transition temperature converges to a fixed point; the final converging temperature agrees with the one determined by the MC $\tilde C$-peak, 2.03. This study suggests that the approximate location of the transition temperature can be already estimated by using early training ranges far away from the transition point and that the more precise determination 
can be achieved by progressively adding configurations closer to the transition point.
The procedure actually
reveals a mechanism of finding a phase transition point, without \emph{a priori} knowledge of its existence, by taking two small training ranges as the starting point and proposing a phase transition point somewhere in between.

To answer the second question, we conduct a numerical experiment on a NN with GSM MC data in the reduced temperature range $[2.5,5.5]$. Within this coil region, both $S^2$ and average energy (not shown) have significant variations. We enforcedly train the NN so that the two output neuron nodes mistakenly regard configurations in the
range $[2.5,3.5]$ as in phase-1 and $[4.5,5.5]$ as in phase-2. We then test the network with independent configurations over the entire $[2.5,5.5]$ range. The results of the output nodes are plotted in Fig. \ref{compare}(b). Each node vaguely recognizes the configurations as ``phase-1'' or ``phase-2'', with a mean $\nu$-value hovering around $0.5$ in high uncertainties. No clear signals, such as those determined above for true phase transitions, exist.

{\it Summary.--} 
We described the training of neural networks to recognize diversely and subtly different polymer states produced from Monte Carlo simulations. One advantage of this approach is directly sending the configuration data represented by molecular coordinates to a NN, \emph{without} defining order parameters or calculating the heat capacity; these are conventionally used in computer simulations to rigorously determine a transition point. In particular, in the low-energy 
regime, a simulated system often encounters potential-energy traps which need to be treated by using a non-Boltzmann weight. Taking the Wang-Landau algorithm as an example, the calculation of the heat capacity in the extremely low-energy regime requires high numerical precision of the computed density of states, which is achievable but requires extensive computations. The NN process describe here, on the other hand, does not require such precision, as long as independent configurations used for supervised training are produced by a numerical simulation. 

The example used here is a classical molecular system displaying gas-, liquid-, and crystal-like structures at various energies.
We demonstrated that NN can classify
both first- (globule to anti-Mackay) and second-order (coil-to-globule and anti-MacKay to Mackay) transitions.
The direct use of molecular coordinates as input into the NN underlies the robustness and simplicity of our approach,
and suggests that other simulation tools, such as molecular dynamics, could be used as well. 
The outcome of this work provides a compelling reason to incorporate machine learning techniques into
molecular simulations more generally, as a powerful hybridized computational tool for the future study of polymeric systems.

{\it Acknowledgement. --} Financial support from the Natural Sciences and Engineering Council of Canada is gratefully acknowledged. This work was made possible by the facilities of the Shared Hierarchical Academic Research Computing Network and Compute Canada.

\bibliographystyle{apsrev4-1}
\bibliography{ref2}


\end{document}